# Memership Inference Attacks Against Latent Factor Model

Dazhi Hu

College of Information Science and Technology, Donghua University, Shanghai China
2222053@mail.dhu.edu.cn

**Abstract.**
The advent of the information age has led to the problems of information overload and unclear demands. As an information filtering system, personalized recommendation systems predict users' behavior and preference for items and improves users' information acquisition efficiency. However, recommendation systems usually use highly sensitive user data for training. In this paper, we use the latent factor model as the recommender to get the list of recommended items, and we representing users from relevant items Compared with the traditional member inference against machine learning classifiers. We construct a multilayer perceptron model with two hidden layers as the attack model to complete the member inference. Moreover,a shadow recommender is established to derive the labeled training data for the attack model. The attack model is trained on the dataset generated by the shadow recommender and tested on the dataset generated by the target recommender. The experimental data show that the AUC index of our attack model can reach 0.857 on the real dataset MovieLens, which shows that the attack model has good performance.

**Keywords:** membership inference, recommender systems, collaborative filtering, LFM.

## 1 Introduction

A recommender system is an information screening system that can predict a user's evaluation or preference for an item. With the advent of the Internet age, human beings have moved from "information shortage" to "information power". Consumers hope to find content they are interested in in the massive information, and it is difficult for information producers to differentiate their products and attract people's attention, which leads to the problem of "information overload". As a tool to link users with information, recommender systems can better deal with the problems of "information overload" and "long tail", so as to provide users with personalized services. There are three common recommendation methods: 1) Content-based recommendation algorithms. 2) Collaborative filtering recommendation algorithm. 3) Hybrid recommendation algorithm.

  However, the success of recommender systems is based on large-scale user data, which is very likely to leak users' privacy, such as users' facial information, social information, location information, medical information, etc. For example, as one of the biological features of the human body, face information, like fingerprints and iris, has become a means of proving one's identity. Face recognition technology, which performs identity verification based on human facial features, has been widely used, such as mobile phone unlocking, registration verification of various APPs, lending and so on. According to security industry sources, using tools such as PS, you can create a face image with a background, and then through dynamic video software, the face photo can realize actions such as blinking, nodding, etc., combined with the now very mature AI face changing Technology, with some audio simulation technology and personal privacy information, will be used by criminals for video chat fraud. This very confusing fraud method is not easy for ordinary people to identify, so it brings serious information security risks. It is true that the wide application of machine learning has brought people a lot of convenience, but it has also brought about the problem of user privacy protection [1,44].

Therefore, this paper attempts to study the member speculation attack mechanism in the implicit semantic recommendation system. The attacker's goal [20] is to determine whether the target recommender uses the user's data for training, and to evaluate the performance of the attack model through indicators such as AUC, so as to gain a more comprehensive understanding of the recommendation system. Therefore, defense measures such as popular randomization are proposed to achieve the goal of privacy protection.

In this paper, a latent semantic recommendation system based on matrix decomposition is used to generate a user-item matrix, and a stochastic gradient descent algorithm is used to minimize the loss function and improve the recommendation performance. In the past, most of the member speculation attacks of machine learning classifiers [30] were at the sample level, while this paper focuses on the user-level member speculation attacks, that is, to determine whether the target recommender uses the user's data for training, based on the user's interaction with the item and the recommendation system. For user's recommendation, we extract the user's feature vector as the input of the attack model. Compared with the previous work of most members speculating on the attack, the attacker can only observe the recommended items from the recommender system, not as the posterior probability. Recommended results. User-level member speculation attack has wider application fields, and it can also make us better understand the mechanism of member speculation attack.

## 2   Recommender system related background knowledge

### 2.1   Recommendation system definition and value

People are looking for what they need in the massive information, and at the same time, the massive information is also looking for the right person. Two important preconditions for the emergence of the system arise, namely information overload and unclear requirements. Recommendation systems are widely used now. The most common example is Taobao's recommendation algorithm, which can recommend things that you may be interested in based on your basic information, such as gender, region, browsing. The recommender system essentially solves the problem of matching users, items and environments, and helps to establish connections between users and items. The recommendation system is a branch application of machine learning. The recommendation system uses a lot of machine learning technology, and uses various algorithms to build recommendation models to improve the accuracy, surprise, coverage, etc. of recommendations, and even try to feedback changes in users' interests, such as Today's Toutiao APP pulls down to display new news, and feedback changes in users' interests in real time. The recommender system well meets the needs of the "target" provider, the platform, and the user. Taking Taobao shopping as an example, the provider of the "subject matter" is Taobao's thousands of store owners, the platform is Taobao, and the user is the natural person or enterprise shopping on Taobao. Through the recommendation system, products can be better exposed to users who need to buy, and the allocation efficiency of social resources can be improved.

### 2.2    Existing Recommender System Algorithms

The recommendation algorithm is the core of the recommendation system. With the continuous development of the recommendation system, the types of recommendation algorithms are also increasing, and more and more attention has been paid by the academic and industrial circles. Existing recommendation algorithms [21]can be divided into content-based recommendation, collaborative filtering recommendation and hybrid recommendation. Collaborative filtering recommendation can be divided into neighborhood-based recommendation and model-based recommendation. In neighborhood-based recommendation, user-based recommendation and item-based recommendation are its two categories, and in model-based recommendation, latent semantics, latent semantic model and graph model are the main research hotspot models [16,40-43].

### 2.3    Latent Semantic Model Recommendation System Algorithm

The recommendation algorithm adopted in this work is the implicit semantic model algorithm. The latent semantic model algorithm was first proposed in the text field to find the hidden semantics of the text. In 2006, it was used for recommendation. Its core idea is to link user interests and items through implicit features, find potential topics and categories based on user behavior, and then automatically cluster items into different categories or theme.

## 3 Member Speculation Attack Design in Recommender Systems

### 3.1 Member Speculation Attack Definition

At present, most machine learning systems are designed with only a weak threat model in mind. When faced with natural input, the system performance can be better reflected, but when encountering malicious users, the performance of these systems may be greatly threatened.

A common definition of a membership inference attack is that, given access to a data record and a target model, an attacker needs to determine whether that record is in the model's training dataset. Specifically, it will directly reveal their private information if it is known that the user's data is part of the recommender system's training data. The essence of the member speculation attack is a binary classifier. The performance of machine learning models on their training data and the test set data they encounter for the first time is often different. The attacker's goal is to build an attack model that can distinguish the behavioral differences of the target model, and use this attack model to identify the target model. members and non-members.

After synthesizing the shadow dataset $D'$, the attacker starts to simulate the target model. If the target model is trained on a machine learning service platform, the attacker can use the same service to train a simulated model. The purpose of this step is to train a simulated model with the same or similar predictive ability as the target model. In this way, the attacker can clearly grasp the information of the training data and test data of the simulated model, thereby providing training data for the attacking model. Based on the obtained attack model data, the attacker uses machine learning models such as Naive Bayes, decision tree, and neural network to train the attack model

### 3.2 Member speculates attacking adversary information

**enemy attack target**

The adversary target means the degree of damage caused by the attacker's target, which can be divided into: (1) Confidentiality attacks. Attackers try to steal structural parameters about the target model, as well as confidential information such as training data that contains sensitive information. (2) Integrity attack. Attackers try to induce model output, affect the integrity of the training process, or affect the output of the model's prediction phase. (3) Availability attacks. Attackers attempt to affect model performance or quality of service, hinder or interrupt normal user requests for the model, and make it untrustworthy in the target environment.

**Adversary Background Information**

The knowledge of the adversary inferred by the members of the attack is divided into three categories: black-box knowledge, gray-box knowledge, and white-box knowledge. Black-box knowledge: When the attacker does not have any expertise about the training data, the attacker has black-box knowledge at this time. Grey-box knowledge: When the attacker knows part of the expertise about the training data, the attacker has grey-box knowledge. White-box knowledge: An attacker with white-box knowledge can obtain a certain version of the real data of the training data, and can train the corresponding mirror model according to this version, or can use all the knowledge of the network structure and parameters to attack the model.

## 4  Evaluation of Member Speculation Attack Performance in Recommender Systems

### 4.1  Experimental setup

**data set**

The data used in this experiment is the MovieLens dataset provided by GroupLens, which is a public dataset. The MovieLens dataset has been widely used as an effective dataset that can be widely used and validated for algorithms. This dataset provides a good source for recommender system scientists and researchers, and many recommendation algorithms have been developed and validated with this dataset. The dataset has been denoised and necessary processing and can be used directly. The dataset used in this paper contains 100,836 ratings data for 9,742 movies by 610 users during the period from 1996.3.29 to 2018.9.24. Users are randomly selected. All selected users have rated at least 20 movies. Demographics are not included, each user is represented by an id, and no other information is provided. Rating scores ranging from 1 to 5 indicate how much the user likes the movie. The dataset is randomly divided into two disjoint subsets, namely the shadow dataset and the target dataset.

**Recommended method**

Recommendation algorithms output recommended items based on the information learned from the input. In this paper, the latent semantic algorithm is adopted, the stochastic gradient descent algorithm is used to update the parameters with a learning rate of 0.01, and the model is constructed with a regularization coefficient of 0.01 to enhance the generalization ability of the model. The adversary can observe the list of recommended items from the recommender system and employ a matrix factorization method to project user movies into a shared latent space.

### 4.2  Experimental results

To evaluate the attack performance, it is assumed that the adversary has a shadow dataset from the same distribution as the target recommender's training data and knows the target recommender's algorithm, for this paper, both the target recommender and the shadow recommender are Use latent semantic models. Experimental results show that our attack model has good performance. Plotting the ROC curve, we get the following image:

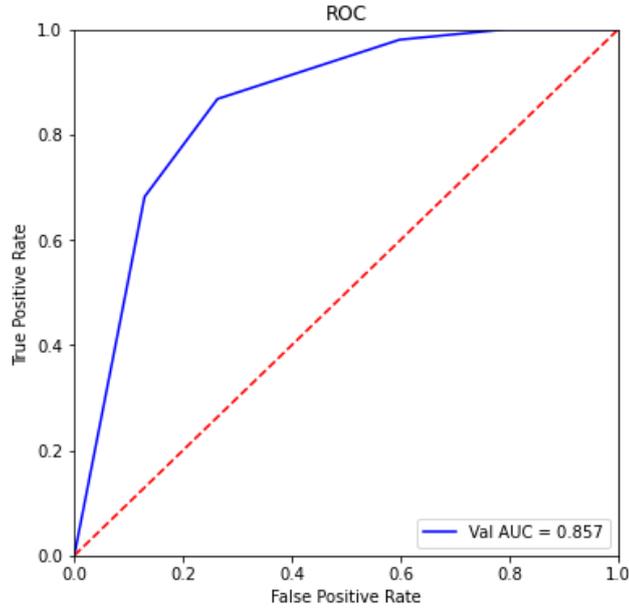

**Fig.1**. MLP model experiment ROC curve

Next, we try to explore the influence of the length k of the vector on the attack performance. We evaluate the attack model performance from 10 to 100 different values, as shown in Figure 4-2, the abscissa is the vector length, and the ordinate is the value of AUC. According to the experimental results, we get the following conclusions: when the vector length is less than 50, the attack performance improves with the increase of k. When the value of k exceeds 50, the attack performance will theoretically be improved, because a larger length vector can provide more dimensional perspectives, however, when the vector length is large enough, the performance of the attack model tends to be stable.

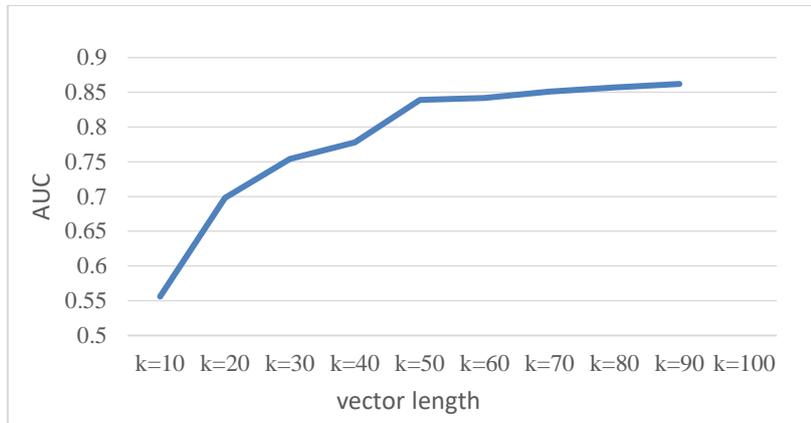

**Fig. 2**. Influence curve of vector length on attack performance

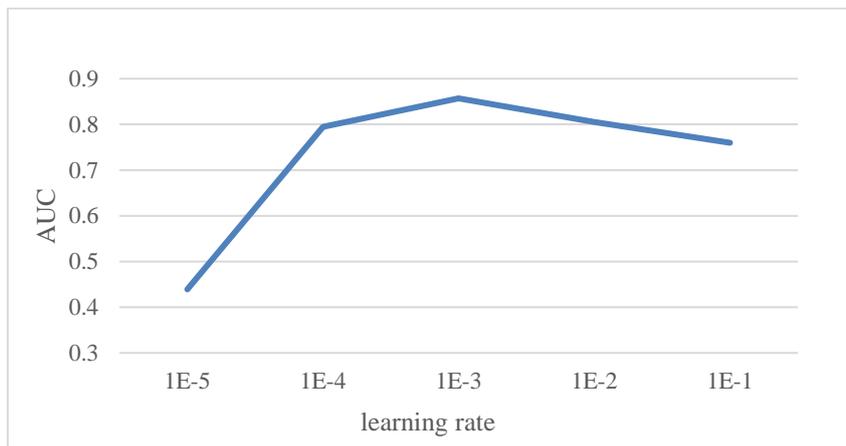

**Fig. 3**. Model attack performance under different learning rates

## 5    Conclusion and Outlook

### 5.1    Conclusion of the work of this paper

Personalized recommender systems have received more and more attention with the deepening of research and the emergence of the needs of the Internet industry. Not only traditional electronic websites need the functions and services of recommender systems, but some emerging fields also introduce recommender systems. Music, movies, news are all recommended objects. This paper introduces several widely used recommendation algorithms and the related work of member speculation attack, introduces its basic principles, and proposes the recommendation algorithm and member speculation attack model used in this experiment. In this paper, a personalized recommendation implicit semantic model is used to generate a recommendation list, and then the adversary extracts the user's feature vector, that is, the difference between the interaction and the center vector of the recommendation set, as the user's feature vector. The user feature vector and labels are then used as input to the attack model, which is trained on the shadow dataset and tested on the target dataset.

The disadvantage of this paper is that the recommendation algorithm is relatively single, and the latent semantic model is used for both the shadow recommender and the target recommender. The attack performance of the attack model when the shadow data user and the target recommender use different recommendation algorithms has not been verified. Therefore, the attack model lacks certain generalization ability. In addition, due to the time relationship, this paper does not test defenses against member speculation attacks, such as defense strategies such as reducing the number of categories, publishing data using differential privacy protection, popular randomized recommendation algorithms, etc., which have been obtained in previous work. Effective tests [19][20].

### 5.2 Outlook for future work

This paper studies the member speculation attack of recommendation system based on latent semantic model. Due to factors such as time and energy, the shadow recommender and target recommender in this paper are the same recommendation algorithm. In future research, the attack performance of different recommendation algorithms can be tested. , such as item-based recommendation, neural network collaborative filtering recommendation algorithm. For example, the shadow recommender uses a neural network collaborative filtering algorithm, while the target recommender uses a latent semantic recommendation algorithm to test the generalization ability of the attack model based on different collocations. Furthermore, the impact of hyperparameters on attack performance can be analyzed. Testing the attack performance from different aspects can effectively improve the attack generalization ability of the model, and also help us to fully understand the attack mechanism of the members to propose a better defense mechanism.